\documentclass[emulateapj,apj]{emulateapj}

\shorttitle{TYPE Ia SUPERNOVA IN THE SDSS}
\shortauthors{HAN ET AL.}

\begin{document}

\title{The Properties of Type I\lowercase{a} Supernova Host Galaxies \\
    from the Sloan Digital Sky Survey}

\author{Du-Hwan Han\altaffilmark{1}, Changbom Park\altaffilmark{2}, Yun-Young Choi\altaffilmark{3, 4}, and Myeong-Gu Park\altaffilmark{1}}

\altaffiltext{1}{Department of Astronomy and Atmospheric Sciences, Kyungpook National University Daegu 702-701, Korea; duegdo13@gmail.com, mgp@knu.ac.kr}
\altaffiltext{2}{Korea Institute for Advanced Study, Hoegiro 87, Dongdaemun-Gu, Seoul 130-722, Korea; cbp@kias.re.kr}
\altaffiltext{3}{Department of Astronomy \& Space Science, Kyung Hee University, Kyungki 446-701, Korea; yychoi@kias.re.kr}
\altaffiltext{4}{Corresponding author}

\begin{abstract}
 We investigate the properties and environments of Type Ia Supernova
(SN Ia) host galaxies in the Stripe 82 of the Sloan Digital Sky Survey-II
Supernova Survey centered on the celestial equator. Host galaxies
are defined as the galaxy nearest to the supernova (SN) in terms of angular
distance whose velocity difference from the SN is less than $1000$
km s$^{-1}$. Eighty seven SN Ia host galaxies are selected from the SDSS
Main galaxy sample with the apparent $r$-band magnitude $ m_r < 17.77$,
and compared with the SDSS Main galaxies.
The SN Ia rates for early and late-type galaxies are
$0.81 \pm 0.19$ SN $(100\;\rm{yr})^{-1}$ and $0.99 \pm 0.21$ SN $(100\;\rm{yr})^{-1}$,
respectively.
We find that the host galaxies have a color distribution consistent with
that of the Main galaxies, regardless of their morphology.
However, host galaxies are on average brighter than the Main galaxies by
$\sim 0.3$ mag over the range of $-18.3 > M_r > -21.3$. But the brighter ends of their
luminosity distributions are similar. The distribution of the distance to
the nearest neighbor galaxy shows that SNe Ia are more likely to occur in
isolated galaxies without close neighbors.
We also find that the SN Ia host galaxies are preferentially located
in a region close to massive galaxy clusters compared to the Main galaxies.

\end{abstract}

\keywords{galaxies: fundamental parameters --- galaxies: statistics --- supernovae: general}

\section{Introduction}
For over a decade Type Ia supernovae (SNe Ia) have widely been used
as a cosmological distance indicator to directly probe the accelerating
expansion of the universe (Riess et al. 1996; Perlmutter
et al. 1997, 1999; Astier et al. 2006) and measure the Hubble
constant $H_0$ (Jha et al. 1999). In addition, supernovae (SNe) are critical
in the study of star formation history and the chemical evolution of the
universe (Sharon et al. 2007). SNe Ia are believed to be the result of
a deflagration of a white dwarf that accretes matter from a companion or
merges with another white dwarf (Whelan \& Iben 1973; Iben \& Tutukov 1984;
Webbink 1984). Yet, the stellar evolution leading to SNe Ia is not clearly
understood.

Recent studies have focused on the correlation between the SNe Ia and their host galaxies,
trying to understand SNe Ia through the analyses of properties of their host galaxies.
The characteristics, such as the morphology, color, star formation rate, metallicity, and
stellar age, of host galaxies provide clues to the understanding of
the progenitors (Gallagher et al. 2005; Mannucci et al. 2005; Sullivan et al. 2006;
Calura \& Matteucci 2006).
Gallagher et al. (2005) studied the effect of the environment on the
properties of SNe Ia by analyzing SN Ia host galaxies to investigate
systematic effects in the calibration of SNe Ia.
Mannucci et al. (2005) studied the dependency of the SN Ia rate per
unit stellar mass on the morphology and the color of the host galaxies.
They found that the SN Ia rate is higher in late-type than in early-type
galaxies and larger in bluer galaxies than in redder galaxies.
Sullivan et al. (2006) showed that the SN Ia rate per unit mass is
proportional to the star formation rate of the host galaxies.
Calura \& Matteucci (2006) studied SNe rate as a function of the Hubble
galaxy type using four different chemical evolution models and found
an increasing trend of the SN Ia rate toward the later Hubble type.
The SN rate itself would provide information on the formation and
the evolution of the galaxies.

The study of SNe in galaxy cluster is also important in understanding
the metallicity evolution and star formation history of the intracluster
medium (Mannucci et al. 2008). Research on the cluster SN rate progresses
vigorously as the detection of SNe in galaxy cluster increases
(Gal-Yam et al. 2003; Sharon et al. 2007).
The SN Ia rate in early-type cluster galaxies is found to be higher
than that in early-type field galaxies (Mannucci et al. 2008).
Carlberg et al. (2008) measured the clustering of the
SN host galaxies relative to the field galaxies using the angular
cross-correlation function of the Supernova Legacy Survey sample. They
found that the SN host galaxies are more correlated with galaxy
clusters than the field galaxies.

However, most studies so far are not free from small number statistics and sample biases.
Recently, the Sloan Digital Sky Survey (SDSS) have been completed, and vast amounts of data
on galaxies and SNe have been released, which enabled statistically meaningful
studies on the galaxy properties and the environmental effects on the galaxy properties
with the SDSS galaxy data (Goto et al. 2003; Balogh et al. 2004; Tanaka et al. 2004;
Blanton et al. 2005; Weinmann et al. 2006; Choi et al. 2007; Park et al. 2007, 2008; Park \& Choi 2009).
The vast amount of the SDSS data makes it possible to construct fair samples
for these studies and to minimize the selection effects. The SN host galaxies
in the SDSS locate in the region called ''redshift desert'' between
the low and high redshift regions, while those used in many of the previous studies
locate in the low and high redshift regions.

The purpose of this study is to understand the nature of the progenitors through the
comparison of the SN Ia host galaxies and the Main galaxies
from the SDSS, in terms of galaxy properties (e.g., morphology, luminosity, and color)
and environmental effects.
In this paper we study the dependence of the SN Ia rate on environments attributed to
the host galaxy, to the nearest neighbor galaxy, and to the nearest cluster of galaxies
(Park \& Choi 2009; Park \& Hwang 2009). The Main galaxies of the SDSS located
in the same environment are used as a comparison sample. The systematic dependence
of the SN Ia rate on host galaxy properties or on the effects of the nearest neighbor
galaxy or cluster will let us understand the nature of the SN Ia progenitors and its
potential redshift dependence.
In Section 2, we describe the SN Ia data from the SDSS-II SN
survey, the Main galaxies, the Abell cluster and identify SN Ia
host galaxies.
The analyses and results on the properties and the environments are
presented in Section 3. We summarize in Section 4.
Throughout this paper we assume a flat universe with density parameters
$\Omega_m = 0.27$, $\Omega_{\Lambda}=0.73$,
and Hubble constant $H_0=100h$ km s$^{-1}$ Mpc$^{-1}$.

\section{Data}

\subsection{The Sloan Digital Sky Survey}

 The SDSS is a large-scale photometric and
spectroscopic survey of Northern Galactic hemisphere using a
dedicated 2.5 m, f/5 survey telescope with a wide field of view
($3^\circ$) at the Apache Point Observatory, New Mexico (York et al.
2000; Gunn et al. 2006). The photometric survey uses a specially
designed multi-band CCD camera that covers five bands over a wide
wavelength range denoted by $u, g, r, i,$ and $z$ with effective wavelengths
of $3550${\AA}, $4770$ {\AA}, $6230${\AA}, $7620${\AA}, and
$9130${\AA}, respectively (Fukugita et al. 1996; Gunn et al. 1998).
The observed objects are processed by an automated photometric pipeline
and are astrometrically calibrated (Lupton et al. 2001). The objects of
SDSS spectroscopic observations such as galaxies, quasars, luminous
red galaxies, and so on are selected by a spectroscopic selection
algorithm (Richards et al. 2002; Eisenstein et al. 2001). The first and second
phases of the SDSS (SDSS$-$I and II) have been completed and produced seven
data releases (see Abazajian et al. 2009 for Data Release 7) so far.

 The SDSS-II Supernova Survey is a part of the SDSS-II project that
scans a designated region called the Stripe 82, centered on the
celestial equator in the Southern Galactic hemisphere, $-60^{\circ} <$
R.A. $< 60^{\circ}$ and $-1^{\circ}_.25 < \delta < 1^{\circ}_.25$,
covering an area of 300 deg$^2$ (Frieman et al. 2008; Zheng et al. 2008;
Sako et al. 2008; Dilday et al. 2008). The SDSS-II SN Survey aims to
discover high-quality light curves for SNe Ia at $0.05
\lesssim z \lesssim 0.35$ and has several advantages such as covering a
larger spatial volume than other SN surveys by using wide-field CCD
camera and drift scanning.
The SDSS-II SN Survey is carried out during three seasons,
September through November of 2005--2007 and found about $1500$
SNe (Frieman et al. 2008).
The public SDSS SNe candidates are listed on the World Wide Web.\footnote{\texttt{http://sdssdp47.fnal.gov/sdsssn/sdsssn.html}}

\subsection{The Main Galaxies}

The SDSS photometric survey covers $\sim$360 million objects over 11,000 
deg$^2$. The photometric catalog derived from the SDSS photometric 
survey lists the photometric parameters, such as apparent magnitude, color, 
photometric redshift, and the extinction, which is also used for the target selection 
of spectroscopic observations (Eisenstein, et al. 2001).

The SDSS Main galaxy sample is derived from the SDSS spectroscopic survey.
Analysis of SDSS Main galaxy spectroscopic sample (Strauss et al. 2002) has
 typically been limited to a galactic extinction corrected Petrosian magnitude 
 $m_r = 14.5$ at the bright end due to two effects: (1) the bright galaxies with large 
 flux within 3$\arcsec$ fiber aperture cause saturation and cross-talk in the 
 spectrographs, and (ii) their large angular size and substructures make them shredded 
 into several objects and then cause problems for the spectroscopic target selection.
The exclusion of the galaxies brighter than $m_r = 14.5$ leads to decrease in the possible 
range of luminosity to be explored at a given redshift and in the overall number of galaxies 
when a volume-limited sample is constructed.

A subset of the SDSS Main galaxy sample used in many studies is the New York 
University-Value Added Galaxy Catalog Large-scale Structure 
Sample\footnote{\texttt{http://sdss.physics.nyu.edu/vagc/lss.html}} 
({\texttt{brvoid0}}) (NYU-VAGC LSS; Blanton et al. 2005) with apparent magnitude in 
the range $10 < m_r \leq 17.6$ and redshift in the range $0.001 < z < 0.5$.
To supplement bright galaxies to this NYU-VAGC LSS, Choi et al. (2007) and Y.-Y. Choi et al. (2010, in preparation) 
constructed the Korea Institute for Advanced Study-Value Added Galaxy Catalog 
(KIAS-VAGC) by supplementing bright galaxies whose physical parameters have 
been compiled from earlier catalogs, such as Updated Zwicky Catalog (UZC; Falco et al. 1999), 
{\it IRAS} Point Source Catalog Redshift Survey (PSCz; Saunders et al. 2000), 2dF Galaxy 
Redshift Survey (2dFGRS; Colless et al. 2001), and Third Reference Catalogue of Bright 
Galaxies (RC3; de Vaucouleurs et al. 1991).
The KIAS-VAGC sample includes 583,946 galaxies with $10 < m_r \leq 17.6$ of 
NYU-VAGC LSS, {\texttt{brvoid0}}, based on SDSS Data Release 7 (DR7) and 
10,497 galaxies with $10 < m_r \leq 17.6$ (1455 with $10 < m_r \leq 14.5$) 
whose redshifts come from sources other than SDSS.
In this work, we extend the apparent magnitude limit to $m_r = 17.77$ at the faint end, 
which add 114,303 redshifts of galaxies with $17.6 < m_r \leq 17.77$ in the LSS 
Sample ({\texttt{full0}}) of NYU VAGC to the KIAS-VAGC above. Therefore, in total we 
use 708,746 galaxies.
This KIAS-VAGC contains 20,167 galaxies distributed in a region with 
$-51^\circ <$ R.A. $< 60^{\circ}$ of Stripe 82. We use these galaxies to construct 
our Main galaxy sample in this study.
The Main galaxies in NYU-VAGC at a distance below 64 $h^{-1}$ Mpc are corrected for 
peculiar velocity using a model of the local velocity field based on {\it IRAS} 1.2 Jy redshift survey 
(Willick et al. 1997). Outer that radius, the peculiar velocity can be neglected 
(see the caveat Blanton et al. 2005 for details).

The morphology of galaxies in the Main galaxy sample is classified by the automatic 
morphology classification of Park \& Choi (2005). It is a new morphology classification 
scheme that uses the color--color gradient space and the concentration index. 
The morphology of the Main galaxies is classified into two types: early (ellipticals 
and lenticulars; E and S0) and late (spirals and irregulars; S and Irr) types.
The completeness and reliability of the morphology classification reaches about 
$90\%$ (Park et al. 2008; see Table 1 of Park \& Choi 2005).

\begin{deluxetable*}{cccc}
\tablewidth{0pt}
\tablecaption{List of SN Ia Host Galaxies and the Main Galaxy Samples \label{Tsample}}
\tablehead{
\colhead{Sample}  & \colhead{$N_{\textmd{host}}$ ($N_{\textmd{early}}$/$N_{\textmd{late}}$)} &
\colhead{Redshift Criteria} & \colhead{Remarks} }
\startdata
Host-1 & 87 (38/49) & All range & SN Ia host galaxies with $m_r <17.77$ \\
Main-1 & 20,167(8,815/11,352)&All range & KIAS-VAGC with $10 < m_r < 17.77 $ \\
Host-2 &21 (7/14) & $z \le 0.08$ & SN Ia host galaxies consistent with \\
& & & criteria of the Abell clusters, $ z \le 0.08$ \\
Main-2  & 6,534(2,260/4,274) & $z \le 0.08$ & Subsample of the Main galaxies consistent with \\
& & & criteria of the Abell clusters, $ z \le 0.08$\\
Host-3 & 77 (32/45) & $ 0.03 \le z \le 0.17$ & Subsample of Host-1 galaxies consistent \\
& & & with criteria of density tracers\\
Main-3  & 16,427 (6,935/9,492) & $0.03 \le z \le 0.17$ & Subsample of the Main galaxies consistent\\
& & & with criteria of density tracers\\
Host-4 & 40 (18/22)    & $0.03 \le z \le 0.14$ & SN Ia host galaxies with $M_r \leq -20.4$ \\
Main-4 & 5,921(2,950/2,971)& $0.03 \le z \le 0.14$ & KIAS-VAGC with $M_r \leq -20.4 $
\enddata
\end{deluxetable*}

\subsection{Type Ia Supernovae and their Host Galaxies}

\begin{figure}
\epsscale{1.2}
\plotone{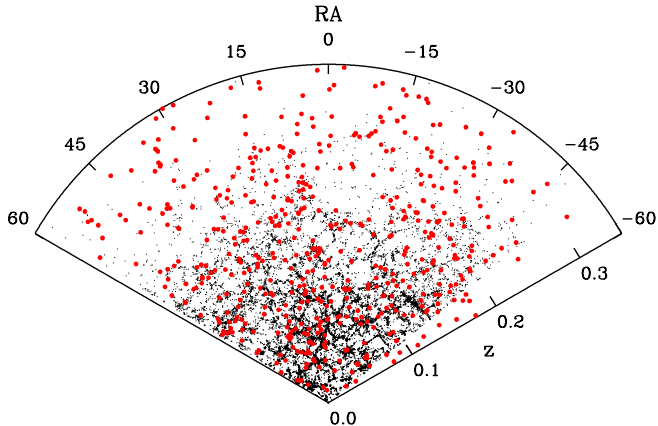}
\caption{Distribution of SNe Ia and the Main galaxies
with  $14.5 < m_r < 17.77$ in the Stripe 82. Dots
are the Main galaxies and filled circles indicate SNe Ia. \label{fig1}}
\end{figure}

During the entire season, the SDSS-II SN Survey discovered about 1,500 
SN candidates, and 612 of them were spectroscopically confirmed 
as SNe Ia. Figure \ref{fig1} shows the distribution of these 612 
SNe Ia in the R.A. versus redshift space. Out of 612 SNe Ia, 594 
SNe Ia are located within the coordinate boundary of the Main galaxy sample 
since the range of the SDSS-II SN Survey is wider in RA than that of the Main 
galaxy sample in the Stripe 82.

We first searched for the host galaxies in the SDSS photometric catalog. A 
galaxy in the catalog that has the smallest, yet no larger than $6\arcsec$, 
angular separation from the SN was identified as the host galaxy. Although 
galaxies in the SDSS photometric catalog have photometric redshifts, their 
errors of 0.024 at 68$\%$ confidence level (Oyaizu et al. 2008) are too 
large to be used for the velocity comparison. We found host galaxies of 512 
SNe Ia in the SDSS photometric catalog. Most of them are faint with the 
apparent magnitude $m_r > 17.77$, therefore not in the spectroscopic Main 
galaxy sample. We could not find the host galaxies of remaining 82 SNe 
either from the photometric catalog or from the SDSS images.

We then searched for the host galaxies of the 594 SNe Ia again in the 
spectroscopic Main galaxy sample. Since all galaxies in the spectroscopic 
sample have accurate redshift information, we can select galaxies that have 
the radial velocity difference from the SNe Ia less than $|\Delta v| = 1000$ km 
s$^{-1}$, out of which we identify the galaxy that has the smallest angular 
separation from the SN as its host galaxy. We found host galaxies of 68 SNe 
Ia this way, all of which were the same host galaxies identified by the 
photometric catalog. So we expect that all 512 host galaxies identified are the 
closest galaxies to the SN in terms of angular separation and within the 
limited volume bounded by the relative velocity difference of 1000 km s$^{-1}$.

Morphologies of the host galaxies from the spectroscopic Main galaxy sample 
were determined by the automatic morphology classification method 
(Park \& Choi 2005) as well as visual inspection method that uses a visual tool 
from the SDSS Catalog Archive Server 
Web site\footnote{\texttt{http://cas.sdss.org/astrodr7/en/}}. Morphologies 
of those from photometric catalog were determined by visual inspection only.

The absolute magnitudes of 512 candidate host galaxies that have the
spectroscopic and photometric data from either the KIAS-VAGC
or the SDSS photometric catalog were calculated.
To determine the absolute magnitude the redshifts of galaxies are needed.
The redshifts of most candidate host
galaxies that are in the SDSS photometric catalog, but not in the Main
galaxy sample, are taken from NASA/IPAC Extragalactic
Database (NED).\footnote{See \texttt{http://nedwww.ipac.caltech.edu/}}
For the remaining host galaxies whose redshifts are not in NED, it was assumed
that the host galaxy has the same redshift as that of the SN.

\begin{figure}
\epsscale{1.2}
\plotone{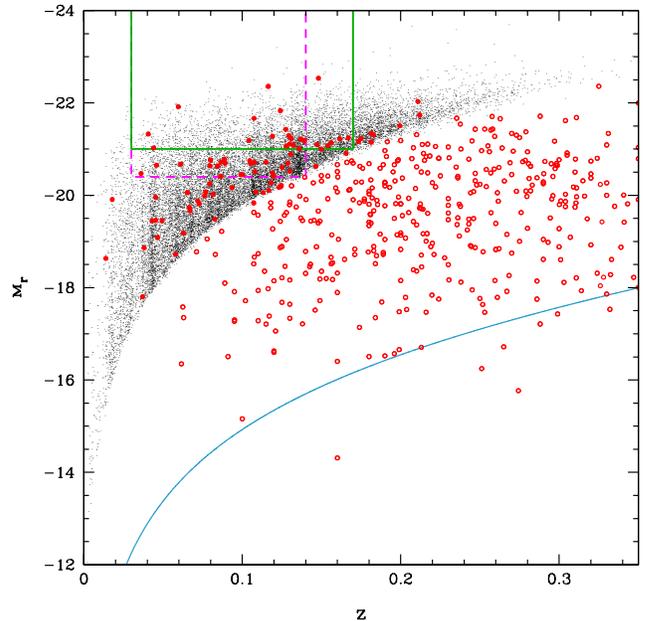}
\caption{Distribution of SN Ia host galaxies and the Main galaxies
in the redshift-absolute magnitude space. Dots are the Main galaxies
and circles indicate SN Ia host galaxies. Filled circles indicate the
Host-1 sample. The bottom solid line represents the apparent
magnitude limit of $m_r=22.5$. The upper solid and dashed rectangular box
represents the definition of the density tracers and volume-limited sample,
respectively. \label{fig2}}
\end{figure}

The absolute magnitudes of host galaxies are calculated from the
following formula
\begin{equation}\label{eq1}
M_r= m_r-5\log[d_c(1+z)]-25-K(z)-\bar{E}(z),
\end{equation}
where $d_c$ is the co--moving distance, $K$ is the $K$-correction
as determined according to Blanton et al. (2003), and $\bar{E}(z)$ is
the mean luminosity evolution correction given by Tegmark et al.
(2004). We adopt a flat universe with matter density $\Omega_m=0.27$ and
cosmological constant $\Omega_{\Lambda}=0.73$ (Spergel et al. 2007).
The co-moving distance is given by
\begin{equation}\label{eq2}
d_c=\frac{c}{H_0}\int_0^z \frac{1}
{\sqrt{\Omega_m(1+z')^3+(1-\Omega_{m})}}dz'
\end{equation}
in the flat universe that we adopt (Park 1996; Fukugita et al. 1992).

\begin{figure}
\epsscale{1.2}
\plotone{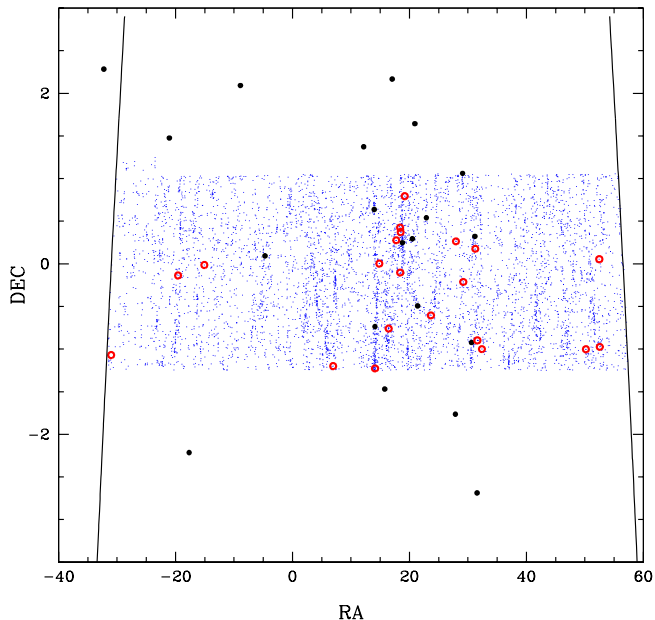}
\caption{Distribution of the Abell clusters and SN Ia host
galaxies. Open circles, filled circles, and
dots indicate the host galaxies, Abell clusters, and the Main galaxies
used in Section 3.2, respectively. The solid line indicates
galactic latitude $b = -40$.
\label{fig3}}
\end{figure}

Figure \ref{fig2} shows the distribution of the SDSS Main galaxies
(dots) in the Stripe 82 and SN Ia host galaxies
(circles) in the redshift versus the absolute magnitude
space. The final SN Ia host galaxy sample consists of 512 galaxies
that are listed in the SDSS photometric catalog and whose redshifts are
obtained either from the SDSS spectroscopic survey or from NED.
Of these host galaxies, 87 host galaxies (Host-1 sample,
filled circles in Figure \ref{fig2}) that satisfy the apparent
magnitude limit of the Main galaxy sample are chosen.

Our sample of 87 host galaxies consists of 38 early-type and 49 late-type
galaxies. Since SDSS-II Supernova Survey for 2005--2007 lasted a total of
nine months and the Stripe 82 includes 8815 early-type and 11,352 late-type
Main galaxies, the naive SN Ia rates for early-type and late-type
galaxies are $0.58 \pm 0.09$ SN $(100\;\rm{yr})^{-1}$ and $0.58 \pm 0.08$
SN $(100\;\rm{yr})^{-1}$, respectively. The errors are those expected from 
the Poisson distribution.

If we consider galaxies within a volume-limited sample that contains the 
most host galaxies, $M_r \leq -20.4$ and $ 0.03 \le z \le 0.14$, 
(the dashed rectangular box in the upper-left corner of Figure 2, see Host-4 and 
Main-4 in Table \ref{Tsample}), the SN Ia rates for early- and late-type 
galaxies are $0.81 \pm 0.19$ SN $(100\;\rm{yr})^{-1}$ 
and $0.99 \pm 0.21$ SN $(100\;\rm{yr})^{-1}$,
respectively. Hence SN Ia events in a late-type galaxy are as frequent as 
or slightly more frequent than in an early-type galaxy. Although we set a 
different volume-limited sample, the rates show similar results.

\subsection{Galaxy Clusters \label{secAbell}}

 We construct a sample of galaxy clusters to study their environmental
effects on the SNe Ia event in galaxies.
From a catalog of 4,073 Abell clusters, whose six richness
classes are defined according to the number of galaxies in the
magnitude interval $m_3$ to $m_3 + 2$ where $m_3$ is the
magnitude of the third brightest members (Abell et al. 1989),
2154 clusters for which redshifts are available in NED were selected.
To guarantee high completeness of the cluster sample,
clusters with the galactic latitude, $|b| > 40^{\circ}$ (Abell 1958; Abell et al. 1989),
richness class 0 and with $z \leq 0.08$ were chosen.
The distribution of the Abell clusters used in this work (filled circles)
and SN Ia host galaxies (open circles) are shown in
Figure \ref{fig3}.
The solid line indicates the galactic latitude $b = -40$.

\section{Results}

Three slightly different samples of SN Ia host galaxies and the Main
galaxies suitable for each analysis are constructed: Host-1 and Main-1
are used for the analysis of the properties and environments of galaxies,
Host-2 and Main-2 for the investigation of the effects by galaxy cluster, and
Host-3 and Main-3 for the analysis of the large-scale environment,
respectively.
The summary of the three samples is listed in Table \ref{Tsample}, and
the details of the samples are described in the following sections.

\begin{figure*}
\epsscale{0.7}
\plotone{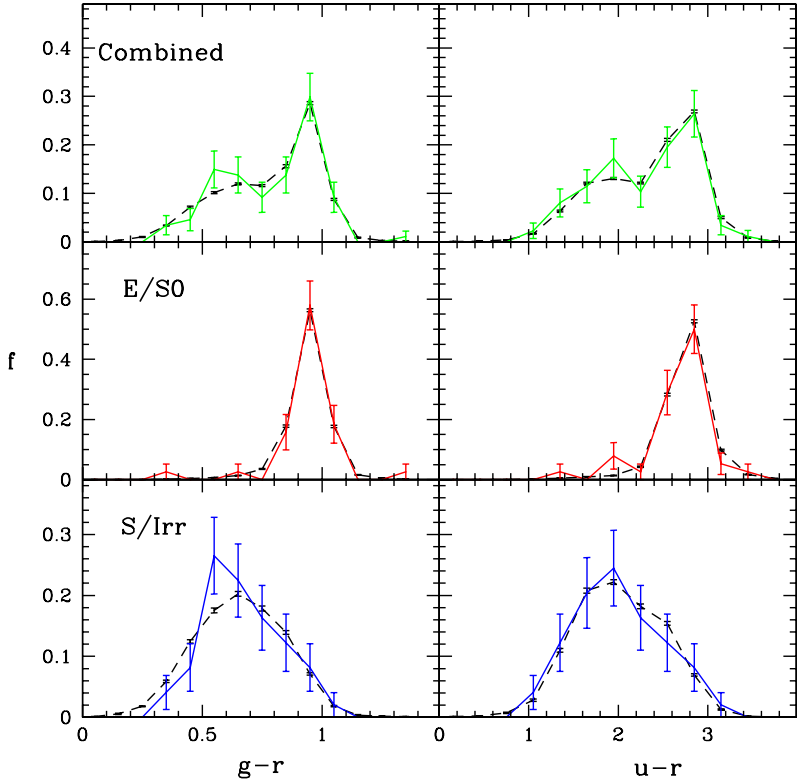} \caption{$g-r$ color (left panels) and $u-r$ color
(right panels) distributions of the Main galaxies (dashed line) and SN Ia
host galaxies (solid line). The top panel is for the combined sample of
the Main galaxies and SN Ia host galaxies. Early types (E/S0) are used in
the middle panel and late types (S/Irr) are used in the bottom panel.
\label{fig4}}
\end{figure*}

\subsection{Properties of SN Ia Host Galaxies}
\subsubsection{Color properties}

To analyze the properties of SN Ia host galaxies and compare them with those
of the Main galaxies, we construct a host galaxy sample (Host-1) that has
the same selection criteria as the Main galaxy sample (Main-1).
We select host galaxies with the apparent magnitude $m_r < 17.77$, the
magnitude limit of the Main galaxy sample, but without the redshift limit.
The absolute magnitude limit corresponding to this apparent magnitude
limit of redshift $z$ is obtained by
\begin{equation}
M_{r,\textrm{lim}}= 17.77 - 5\log[d_c(1+z)]-25-\bar{K}(z)-\bar{E}(z),
\end{equation}
where $\bar{K}(z)$ is the mean $K$-correction given by Choi et al. (2009).
We define a Host-1 sample that consists of 87 host galaxies with $M_r <
M_{r,\textrm{lim}}$. Among 87 host galaxies, 68 are found in the Main galaxy
sample but 19 are found only in the SDSS photometric catalog.
In our morphology classification, 38 are early-type and 49 late-type galaxies.
We construct a comparison sample, Main-1, drawn from the SDSS Main
galaxy catalog. It contains 8,815 early-type and 11,352 late-type galaxies
that are brighter than $m_r =17.77$ and located within the Stripe 82.

\begin{deluxetable*}{ccccc}
\tablewidth{0pt}
\tablecaption{Median of the SN Ia Host Galaxies and the Probability.\label{Tmedian}}
\tablehead{
\colhead{Properties and Environments} & \colhead{Morphology}  &
\colhead{Median of SN Ia Host}  & \colhead{Median of Main Galaxies}  &
\colhead{Probability (\%)}}
\startdata
                    & Combined & -20.61 & -20.31 & 0.8 \\
Absolute magnitude  & E/S0     & -20.64 & -20.47 & 13.4 \\
                    & S/Irr    & -20.45 & -20.19 & 5.6 \\ \hline
                     & Combined & 0.028 & 0.034 & 16.9 \\
Local density        & E/S0     & 0.028 & 0.044 & 12.7 \\
($(h^{-1}$ Mpc$)^{-2}$) & S/Irr    & 0.023 & 0.029 & 23.4 \\ \hline
                                  & Combined & 2.52 & 1.72 & 2.6   \\
Distance to the nearest neighbor & E/S0     & 2.25 & 1.45 & 6.7   \\
    ($r_1/r_{\textrm{vir,nei}}$)          & S/Irr    & 2.70 & 1.92 & 7.6   \\ \hline
Distance to the nearest Abell cluster & Combined & 6.15 & 17.75 & 0.4\\
    ($h^{-1}$ Mpc)  & & & &
\enddata
\end{deluxetable*}

 Figure \ref{fig4} shows the $g-r$ (left panels) and
$u-r$ (right panels) color distributions. The dashed lines
show the color distribution of all (top panels), E/S0-type
(middle panels), and S/Irr-type (bottom panels)
Main galaxies. The solid lines show the color distribution for
each type of galaxies in Host-1 and ''f' means a fraction of galaxies at 
each bin compared to the whole galaxy sample. The figures show that 
the color distributions of host galaxies are similar to that of the Main 
galaxies.

The deviation in the distribution of f and the statistical significance of 
the test for each property and environment are estimated by the 
bootstrap resampling method. The bootstrap resampling infers the 
statistics of samples based on many resamples generated from the 
original sample. Each resample has the same size as the original sample.
For example, to estimate the uncertainty of the host galaxy $g-r$ color 
distribution, we extract each mock galaxy randomly from the host galaxy 
sample until the size of the resample, i.e., the collection of mock galaxies 
generated, reach that of the host galaxy sample.
We generated $10,000$ resamples in this way and calculated the standard 
deviation at each bin from these resamples. The standard deviations are 
shown in Figure \ref{fig4} as error bars.

The statistical significance of the difference between the color
distributions of the host galaxies and the Main galaxies was tested by
comparing the median colors of the host galaxies with that of the Main galaxies.
We generate $10,000$ mock samples from the Main galaxies by the bootstrap
resampling method that had the same sample size as the host galaxy sample.
Then the number of samples whose median color is larger than the
median of the host galaxy sample was counted.

For the $u-r$ and $g-r$ color properties, no significant difference was found
between the medians of the host galaxies and the Main galaxies at $95\%$
confidence level.
However, it is more interesting to investigate the color properties of
SN Ia host galaxies normalized to the stellar masses.

\begin{figure}
\epsscale{1.1}
\plotone{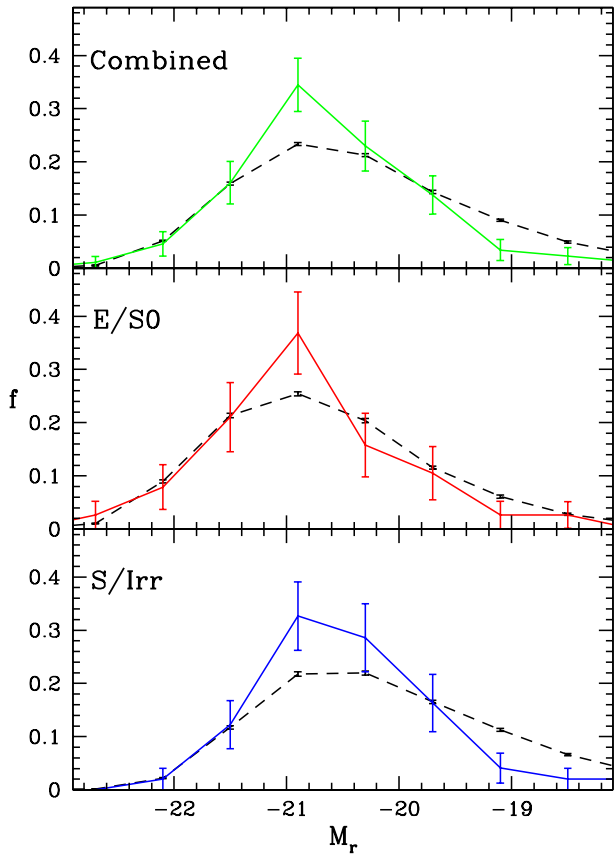}
\caption{Distribution of the $r$-band absolute magnitude of the Main
galaxies (dashed line) and SN Ia host galaxies (Host-1, solid line).
The top panel shows a combined sample of the Main
galaxies and SN Ia host galaxies. Early types (E/S0) are used in the
middle panel and late types (S/Irr) are used in the bottom panel.
\label{fig5}}
\end{figure}

We also checked the SN Ia rate per unit stellar mass.
The stellar masses of SN Ia host galaxies are estimated from the relation
between the colors of galaxies and their stellar mass-to-light ratios. 
Kauffmann et al. (2003) calculated the stellar mass-to-light ratios for about 
120,000 galaxies from the SDSS, which show a tight relation between 
the $g$-band stellar mass-to-light ratio and $g-r$ color 
(Kauffmann et al. 2003). The stellar masses of our host galaxies are 
estimated from this correlation. We find that the SN Ia rate 
per $100 \rm{yr}$ per $10^{11} M_{\odot}$ in the host galaxies bluer than 
$g-r=0.5$, is $0.122^{+0.087}_{-0.077}$, larger than the 
rate $0.008^{+0.005}_{-0.004}$ in the host galaxies with $g-r > 1.0$.
The errors include the standard deviation for the relation of
stellar mass-to-light ratio versus color and the Poisson errors.
Mannucci et al. (2005) reported that the SN Ia rate per
unit mass is higher in the bluer galaxies.

\subsubsection{Luminosity}

In theanalysis of the absolute magnitude, we use the SN Ia host galaxies 
(Host-1) and the Main galaxies (Main-1).
 The absolute magnitude of the host galaxies and the Main galaxies in
the $r$ band are obtained with Equation (\ref{eq1}). Figure \ref{fig5} shows 
the distribution of the $r$-band absolute magnitude, $M_r$, of the host 
galaxies (Host-1, solid line) and the Main galaxies (Main-1, dashed line) 
for each type.
Compared to the Main galaxies, the host galaxies show a higher
fraction near $M_r\sim-21$ which is a little brighter than the
characteristic absolute magnitude $M_{r,\star} \approx -20.3$ of the
Main galaxies (Choi et al. 2007).
Error bars are calculated by the same bootstrap method described in Section 3.1.1.
The distribution of the host galaxies at the luminous end of $M_r<-21.3$
is similar to that of the Main galaxies.
However, the fraction of the host galaxies at the faint end of $M_r > -20$
is lower than the fraction of the Main galaxies while the fraction of the
host galaxies with the intermediate luminosity, $-21.5 < M_r < -20$, is
higher than that of the Main galaxies.
The difference is greater for late-type galaxies as shown in Figure \ref{fig5}.
So it seems that SNe Ia prefer slightly brighter galaxies as hosts,
but not too bright ones.
Gallagher et al. (2005) reached a similar conclusion, and in addition 
showed that the distribution is reasonably well explained by the product 
of the luminosity function of galaxies with the number of stars 
in each galaxy.

We check the median for 74 host galaxies with $-18.3 > M_r > -21.3$.
The median $M_r$ of host galaxies is $-20.61$ while that of the Main
galaxies is $-20.31$. A bootstrap resampling shows that the probability 
$P$ to find by chance a median absolute magnitude of the Main galaxies 
brighter than that of the host galaxies is only $0.8\%$ for the combined 
sample. This means that the median $M_r$ of the host galaxies is 
significantly brighter than that of the Main galaxies at more than $99\%$ 
confidence level.
On the other hand, the median $M_r$ of early-type host galaxies is 
$-20.64$ compared to $-20.45$ of early-type Main galaxies within the same 
magnitude interval. The corresponding probability is $P=13.4\%$. 
The median $M_r$ of late-type host is $-20.45$ compared to $-20.19$ of 
late-type Main galaxies with the corresponding probability of $P=5.6\%$.
The results are summarized in Table \ref{Tmedian}.

\subsection{Environments of SN Ia Host Galaxies}

\begin{figure}
\epsscale{1.1}
\plotone{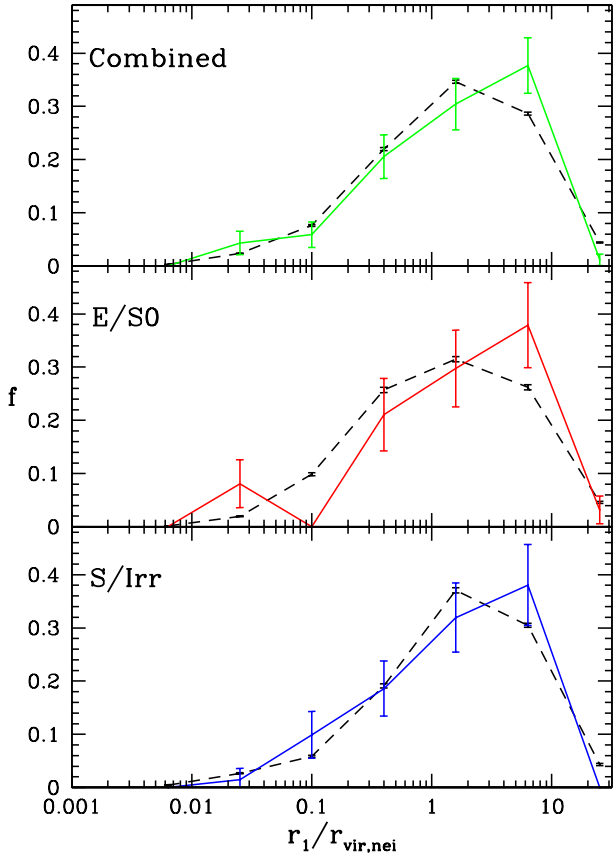}
\caption{Distribution of the nearest neighbor distance of the
Main galaxies (dashed line) and SN Ia host galaxies (solid line).
The distance is normalized by the virial radius of the neighbor galaxy.
Early types (E/S0) are used in the middle panel and late types
(S/Irr) are used in the bottom panel.
\label{fig6}}
\end{figure}

\subsubsection{The nearest neighbor\label{secRnei}}
 We use the distance to the nearest neighbor galaxy as one of the
environmental parameters. Park et al. (2008) and Park \& Choi (2009)
found that the galaxy luminosity and morphology depend on the distance
to the nearest neighbor galaxy and also on neighbor morphology.
In this section, we investigate whether or not occurrence of SNe Ia is
related to the nearest neighbor galaxy.

For both SN Ia host galaxies (Host-1) and the Main galaxies (Main-1),
the nearest neighbor was searched among the Main galaxies and determined
in the following way.
The nearest neighbor is required to have the radial velocity
difference less than $|\Delta v| = 1000$ km s$^{-1}$ with respect to
a target galaxy, and has the smallest angular separation from the target.
Since Park \& Choi (2009) have shown that the important length scale in the
study of galaxy environments is the virial radius of the nearest neighbor
galaxy, the distance was normalized by the virial radius $r_{\textrm{vir}}$ of the
nearest neighbor, defined by
\begin{equation}
r_{\textrm{vir}}=\left(\frac{3\gamma L}{4\pi \times 200\rho_{c}}\right)^{1/3},
\end{equation}
where $\gamma$ is the mass-to-light ratio, $L$ is the $r$-band
luminosity, and $\rho_c$ is the critical density of the universe
(Park \& Choi 2009). Following Park
et al. (2008), we assume $\gamma$(early) $=$ $2\gamma$(late) at the
same $r$-band luminosity, and adopt the mean mass density of the universe
$\bar{\rho}=(0.0223 \pm 0.0005)(\gamma L)_{-20}$ $(h^{-1}$ Mpc$)^{-3}$, 
where $(\gamma L)_{-20}$ is the mass of a late-type galaxy with $M_r=-20$
(Park et al. 2008).

Park \& Choi (2009) found that isolated galaxies are brighter than
those that have neighbors at relatively closer distances.
Hence the dependence of the SN rate on the nearest neighbor distance is
likely to be affected by the dependence on the luminosity of the host galaxy.
To isolate the dependence on the nearest neighbor distance, weights were 
given to host galaxies when the distribution of $r_1/r_{\textrm{vir,nei}}$ is 
calculated.
The weight of a host galaxy with $M_r$ is given by the ratio of the 
fractions of the host and Main galaxies at a given magnitude $M_r$,
i.e., $f_{\textrm{Main}}/f_{\textrm{host}}$ from Figure \ref{fig5}, to remove the
dependency on $M_r$.

Figure \ref{fig6} shows the resulting distribution of the nearest neighbor
distance ($r_1$) normalized by the virial radius of the nearest neighbor
($r_{\textrm{vir,nei}}$) for Host-1 (solid line) and Main-1 (dashed line) galaxies.
It can be seen that the distribution of $r_1$ of SN Ia host galaxies is shifted
to larger separations compared to that of the Main galaxies. This is 
observed for both morphological subsets.

\begin{figure}
\epsscale{1.2}
\plotone{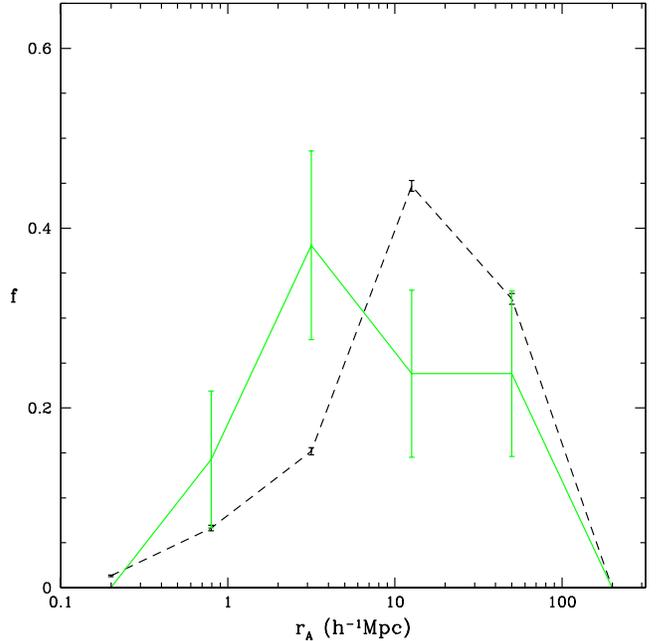}
\caption{Distribution of the distance to the nearest Abell cluster $r_A$.
The dashed line shows $r_A$ of the Main galaxies and the solid line shows that
of the SN Ia host galaxies.
\label{fig7}}
\end{figure}

 The median $r_1/r_{\textrm{vir,nei}}$ of the host galaxies was compared with those
of mock samples drawn from the Main galaxy catalog.
The host galaxies have the median $r_1/r_{\textrm{vir,nei}}=2.52$ compare
to $1.72$ of the Main galaxies. A bootstrap resampling experiment shows
$P=2.6\%$ to have such a difference in the median by chance.
In the case of early-type galaxies we find $r_1/r_{\textrm{vir,nei}}=2.25$ and
$P=6.7\%$, and for late types $r_1/r_{\textrm{vir,nei}}=2.70$ and $P=7.6\%$.
This implies that SNe Ia are more likely to occur in isolated galaxies
without close neighbors within the virial radius.

\subsubsection{Distance to the nearest cluster}

Next, we ask the question if SN Ia host galaxies are in a different cluster 
environment compared to the Main galaxies. Since our selected Abell clusters 
have $z \leq 0.08$, we construct a sample, Host-2, that consists of 21 SN Ia 
host galaxies that are in the Main galaxy sample, therefore $m_r < 17.77$, 
and have $z \leq 0.08$.

The distance to the nearest Abell cluster was derived in two steps.
The velocity difference and the angular separation between
the target host galaxy and the center of a selected Abell cluster was
calculated.
Among the Abell clusters that have a velocity difference smaller than
$\sim 1500$ km s$^{-1}$ with respect to the target galaxy, we select the
nearest one having the smallest projected distance $r_A$ to the
target SN Ia host galaxy at the host galaxy's redshift.
If the projected distance $r_A$ is greater than $15 h^{-1}$ Mpc, then this
host galaxy is not likely to be a member of the nearest Abell cluster.
In such cases,
we choose the three dimensional distance from the center of the
cluster to the target host galaxy with radial distance estimated from
the redshift difference.
Figure \ref{fig7} shows the distributions of $r_A$ of the Main galaxies 
(dashed line) and the host galaxies (solid line).
Two distributions show a visible difference despite large uncertainties.

The Host-2 galaxy sample has the median $r_A$ of $6.14$ $h^{-1}$ Mpc while
the Main galaxies have the median $r_A$ of $17.75$ $h^{-1}$ Mpc, and the
probability of having such a difference in $r_A$ is $P=0.4\%$.
Therefore, it is statistically significant that the host galaxies
are located closer to the Abell clusters compared to the Main galaxies.
Mannucci et al. (2008) reported that cluster early-type galaxies have a
significantly higher SN Ia rate than field early-type galaxies that cannot
be explained by observational biases.
Although our samples are not limited to early-type galaxies,
our result also shows SN Ia host galaxies tend to be located near massive
galaxy clusters.

\subsubsection{Local density\label{secLD}}

\begin{figure}
\epsscale{1.1}
\plotone{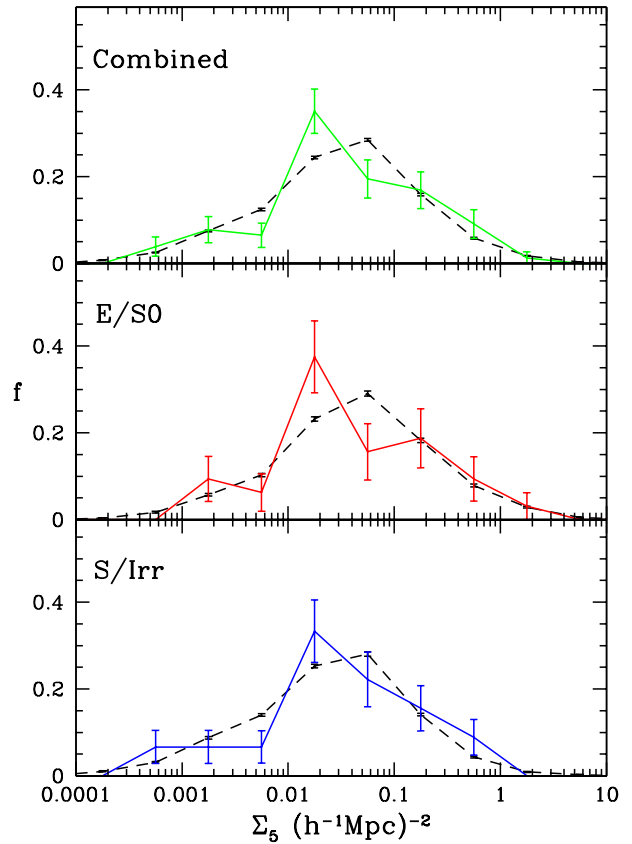}
\caption{Same as in Figure 6 but for the local density. \label{fig8}}
\end{figure}

Many studies have investigated the relationships between the galaxy
properties and the local density (Dressler 1980; Lewis et al. 2002;
Goto et al. 2003; Balogh et al. 2004; Baldry et al. 2006; Cooper et al.
2006, 2007, 2008; Park et al. 2007; Park et al. 2008; Park
\& Choi 2009; Park \& Hwang 2009; Hwang \& Park 2009; Bamford et al.
2009). Dressler (1980) first employed the projected local density
based on counting the number of the nearby neighbor galaxies and defined
the local density as $\Sigma_n = N/(\pi D_{n}^2)$ where
$D_n$ is the distance to the {\it n}th nearest neighbor galaxy.

In this study, we adopt the projected local density based on the fifth
nearest neighbor galaxy as the density estimator.
To search for the fifth nearest neighbor galaxy, a volume-limited sample
of density tracer galaxies was constructed with the absolute magnitude
in $r$-band $M_r \leq -21$ and the redshift in the range $0.03 \le z \le 0.17$
since the redshift limit corresponding to $M_{r,\textrm{lim}} = -21$ is $z=0.17$
in the case of the Main galaxies. The density tracers are those within
the rectangular box at the upper left corner of Figure \ref{fig2}.
A sample of galaxies that are within the same redshift limits
of the density tracer and the apparent magnitude limit of the Main galaxies
were then selected.
We call this sample Host-3. It consists of 31 early types and 46 late types.
We also call the Main galaxies in the same volume as Main-3.

We find the fifth nearest neighbors of the Host-3 and Main-3 galaxies
among the density tracer galaxies.
It is the fifth nearest galaxy in terms of angular separation and has
the velocity difference $|\Delta v| \leq 1000$ km s$^{-1}$ with respect
to a target galaxy.
The adopted local density estimator is
\begin{equation}
\Sigma_5=\frac{5}{\pi D_5^2},
\end{equation}
where $D_5$ is the projected distance from the target galaxy to the
fifth nearest neighbor galaxy at the redshift of the target galaxy.
The distribution of $\Sigma_5$ around SN Ia host galaxies
and the Main galaxies is shown for each morphological subset
in Figure \ref{fig8}.
The solid line and dashed line correspond to the host galaxy sample 
and the Main galaxy sample, respectively.

Figure \ref{fig8} shows that the distribution of $\Sigma_5$ of SN Ia hosts 
has a peak shifted to lower density regions compared to that of the Main 
galaxies.
This is consistent with the finding in Section \ref{secRnei} that the SN Ia 
host galaxies tend to be isolated ones.
There exists a slight excess in SN Ia occurrence for galaxies at very high 
densities ($\Sigma_5 > 0.2$ $(h^{-1}$ Mpc$)^{-2}$), which is again
consistent with the result of the cluster environment study, 
but the statistical significance is not high.
The median $\Sigma_5$ of these host galaxies is $\Sigma_5=0.028$ while 
that of the Main galaxies is $0.035$, and the probability of having such a 
difference is $P=16.92\%$. In the case of early-type galaxies 
we find $P=12.65\%$, and for late types $P=23.40\%$.

\section{Summary  and Discussion}

The SDSS-II Supernova Survey has found about 612 spectroscopically 
confirmed SNe Ia in the Stripe 82 centered on the celestial
equator covering an area of 300 deg$^2$. We found the host galaxies
associated with 512 SNe Ia using the SDSS photometric and spectroscopic
catalogs. The Main galaxies were used as a comparison sample.
A sample of Abell cluster was also used to investigate the effects of nearby
clusters on SN Ia occurrence.
Statistical tests were performed to compare the distribution of the host
galaxies with those of the mock samples generated from
the distribution of the Main galaxies. Our results are as follows.
\begin{enumerate}
\item The color distribution of SN Ia host galaxies is almost same
as that of the Main galaxies, regardless of the galaxy morphology.
However, the SN Ia rate per unit stellar mass is higher in bluer galaxies
than in redder galaxies.
\item Independent of morphological type, we find that SNe Ia are most likely
to occur in galaxies with $M_r\sim-21$ mag. Furthermore, we find an
underabundance of SNe Ia in galaxies brighter than $M_r\sim-19.8$, also
independent of morphological type. Finally, for galaxies brighter than
$M_r\leq-21.5$ the luminosity distribution of SN Ia host is statistically
similar to that of the Main galaxies.
\item Isolated galaxies tend to have SN Ia more frequently than galaxies
having neighbors within their virial radius.
\item SN Ia host galaxies tend to be located near massive galaxy clusters 
in agreement with previous works (Mannucci et al. 2008; Carlberg et al. 2008).
\end{enumerate}

Many studies have been made on SN rate versus
properties and environments of their host galaxies. Cappellaro et al.
(1993) measured the SN rate from 54 SNe discovered by
SN searches carried out at the Asiago and Sternberg Observatories. These
samples have lower redshifts than SDSS samples and contain only 15 SN Ia.
They found that the SN events are proportional to the galaxy luminosity regardless
of the morphology and the SN Ia rate in early types is smaller than that in late
types. This result, however, is uncertain because of the small number of 
sample galaxies.
Our result using a larger sample of 87 SN Ia data suggests that SN Ia events
have dependence on the luminosity of their host galaxies regardless of the
morphology except for most luminous galaxies. The result shows that SNe Ia 
prefer not to occur in too bright or too dim galaxies. However, part of this 
dependence can be explained by selection effect that there are fewer SNe in 
low-luminosity galaxies and SNe hosted by early-type galaxies are fainter 
(Gallagher et al. 2005).

Mannucci et al. (2005) have studied the SN Ia rate per unit stellar mass as
a function of $B - K$ color for nearby galaxies.
They found that blue galaxies have a higher SN Ia rate than red galaxies.
When we normalize the SN Ia rate to the stellar mass, the SN Ia rate shows a
similar result that the SN Ia rate in the bluer galaxies is larger than
in the redder galaxies with respect to $g-r \sim 0.7$.

Sharon et al. (2007) studied cluster SN rate in low-redshift galaxy
clusters with $0.06 < z < 0.19$ and found that the SN rate in a galaxy cluster
is similar to the rate in a field elliptical galaxy. This is also based on 
a small SNe Ia sample.
Our analysis of the environment effects of nearby Abell clusters on the 
host galaxies shows that SN Ia events are more frequent in the galaxies 
located close to massive galaxy clusters. This is in agreement with the 
finding that the host galaxies are more clustered than the field galaxies 
(Carlberg et al. 2008).

A recent study using SDSS-II SN data with $0.05 < z < 0.15$
by Cooper et al. (2009) found that SN Ia events in blue host 
galaxies prefer to occur in the low-density region while there is no 
difference between the environment distributions of red host galaxies and 
galaxies of like properties.
We also find a weak signal that the SN Ia rate is higher for galaxies 
located in relatively low density regions regardless of their morphologies.
On the other hand, analysis of the distance to the neighbor galaxy shows that
SN Ia events are more frequent in isolated galaxies without close neighbors.

\acknowledgments

D.-H.H., C.B.P., and M.- G.P. acknowledges the support of the National Research Foundation of Korea 
(NRF) grant funded by the Korea government MEST (No. 2009-0062868).
Y.Y.C. was supported by a grant from Kyung Hee University in 2010 
(KHU-20100179).

Funding for the SDSS and SDSS-II has been provided by the
Alfred P. Sloan Foundation, the Participating Institutions, the National
Science Foundation, the U.S. Department of Energy, the
National Aeronautics and Space Administration, the Japanese
Monbukagakusho, the Max Planck Society, and the Higher
Education Funding Council for England. The SDSS Web site is
http://www.sdss.org/.

The SDSS is managed by the Astrophysical Research Consortium
for the Participating Institutions. The Participating Institutions
are the American Museum of Natural History, Astrophysical Institute
Potsdam, University of Basel, Cambridge University, Case
Western Reserve University, University of Chicago, Drexel
University, Fermilab, the Institute for Advanced Study, the Japan
Participation Group, Johns Hopkins University, the Joint Institute
for Nuclear Astrophysics, the Kavli Institute for Particle
Astrophysics and Cosmology, the Korean Scientist Group, the
Chinese Academy of Sciences (LAMOST), Los Alamos National
Laboratory, the Max-Planck-Institut f\"{u}r Astronomie (MPIA), the
Max-Planck-Institut f\"{u}r Astrophysik (MPA), New Mexico State
University, Ohio State University, University of Pittsburgh, University
of Portsmouth, Princeton University, the United States
Naval Observatory, and the University of Washington.

\clearpage

\clearpage

\end{document}